\documentclass{PoS}

\title{Latest flow results from PHENIX at RHIC}

\ShortTitle{Latest flow results from PHENIX at RHIC}

\author{\speaker{Eric Richardson}%
        \\
       for the PHENIX Collaboration\\
       University of Maryland, College Park, Maryland, 20742, USA\\
       E-mail: \email{eric.richardson99@gmail.com}}

\abstract{At the Relativistic Heavy Ion Collider (RHIC), key insights into the bulk properties of the hot and dense partonic matter arise from the study of azimuthal anisotropy ($v_2$) of the produced particles.  The $v_2$ values indicate that the matter undergoes rapid thermalization and behaves hydrodynamically at low $p_T$.  Furthermore, the quark scaling of $v_2$ for different particle species suggests that thermalization occurs at the quark level and that $v_2$ is the same for all quark flavors.  Recently, higher order harmonic measurements ($v_3$, $v_4$) have shown the potential for insights into the medium's initial geometry and fluctuations.  This proceeding discusses some of the PHENIX Collaboration's latest flow results and their implications.}

\FullConference{Sixth International Conference on Quarks and Nuclear Physics\\
		 April 16-20, 2012\\
		 Ecole Polytechnique, Palaiseau,  Paris}

\begin{document}

\section{Introduction}

Azimuthal anisotropy in heavy-ion collisions, termed \emph{flow}, is the asymmetric distribution of produced particles with respect to the reaction plane angle ($\Psi_{RP}$), which is defined by the beam axis and the colliding nuclei's impact parameter.  The particle asymmetry distribution is described by the Fourier expansion~\cite{bib:flow_meth}
\begin{equation}\label{eq:fourier_expansion}
\frac{dN}{d(\phi - \Psi_{RP})} \sim \left(1+\sum_{n} 2v_{n}\cos[n(\phi - \Psi_{RP} )] \right),
\end{equation}
where $N$ is the number of particles measured, $\phi$ is the particle's azimuthal angle, $n$ represents the $n^{th}$ harmonic of the particle distribution, and $v_{n}$ is the anisotropy parameter representing the magnitude of the particle asymmetry with respect to $\Psi_{RP}$.  However, due mainly to finite particle statistics and detector granularity, it is impossible to know the reaction plane angle with absolute certainty, thus its experimental measurement is referred to as the event plane angle.

Flow measurements have resulted in many key insights into the created medium.  In particular, the $2^{nd}$ harmonic flow signal, $v_2$, has revealed that the medium undergoes rapid thermalization and behaves hydrodynamically at low transverse momentum ($p_T$).  PHENIX's latest flow results expand upon this knowledge, providing further insight into the medium's properties and initial geometry.

\section{Identified Particle $v_2$}

\begin{figure}
	\centering
  \includegraphics[width=0.80\textwidth]{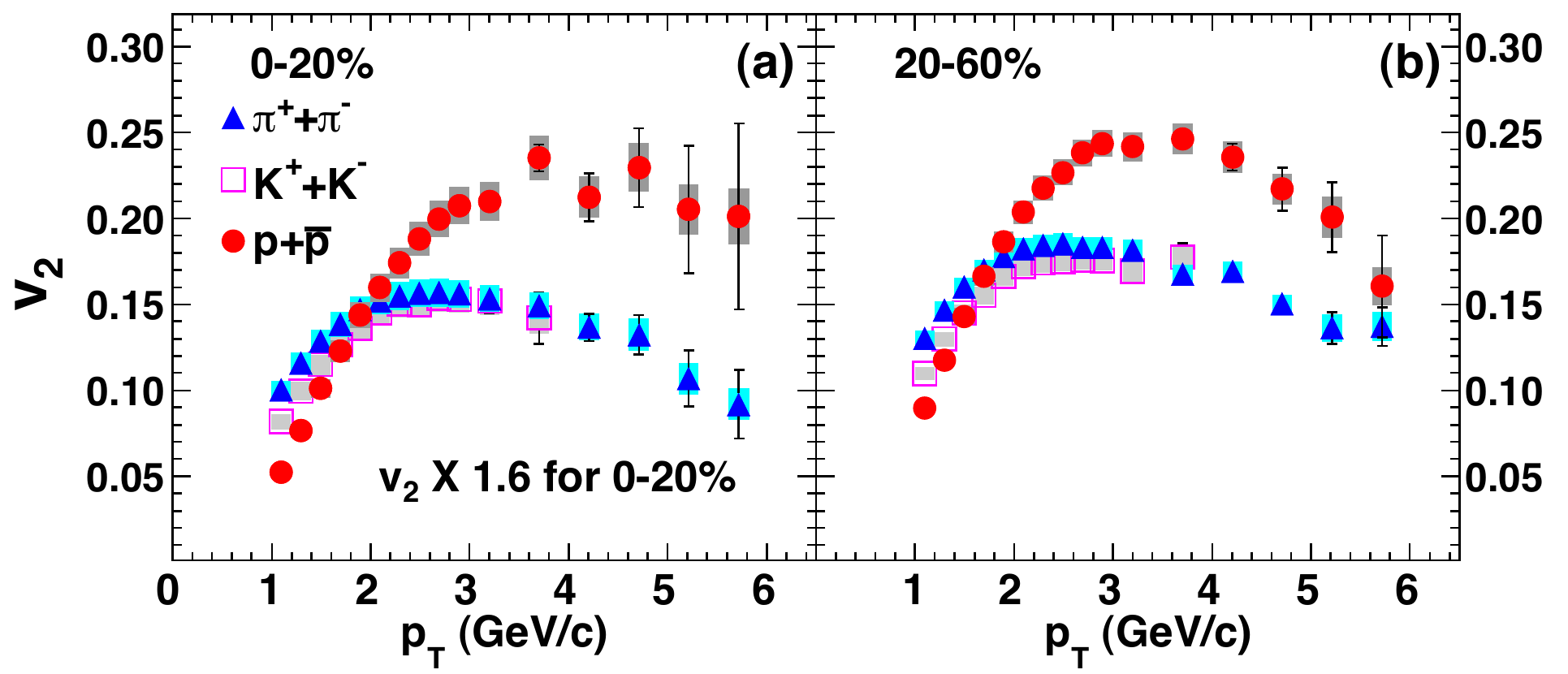}
  \caption{
  	\label{fig:v2_pT_pid}
  	PHENIX identified particle $v_2(p_T)$ from $\sqrt{s_{NN}}$ = 200 GeV Au+Au collisions for centralities of ($a$) 0-20\% and ($b$) 20-60\%~\cite{bib:v2_pid}.  Pions (triangles) and protons (circles) are measured between 1.0 GeV/$c$ $<p_T<$ 6 GeV/$c$ and kaons (squares) 1.0 GeV/$c$ $<p_T<$ 4 GeV/$c$.  Statistical and systematic errors are shown by the bars and boxes, respectively.  Notice in the 0-20\% data the proton signal appears to flatten at high $p_T$, which is not observed for the 20-60\% data. }
\end{figure}

\begin{figure}
	\centering
  \includegraphics[width=0.80\textwidth]{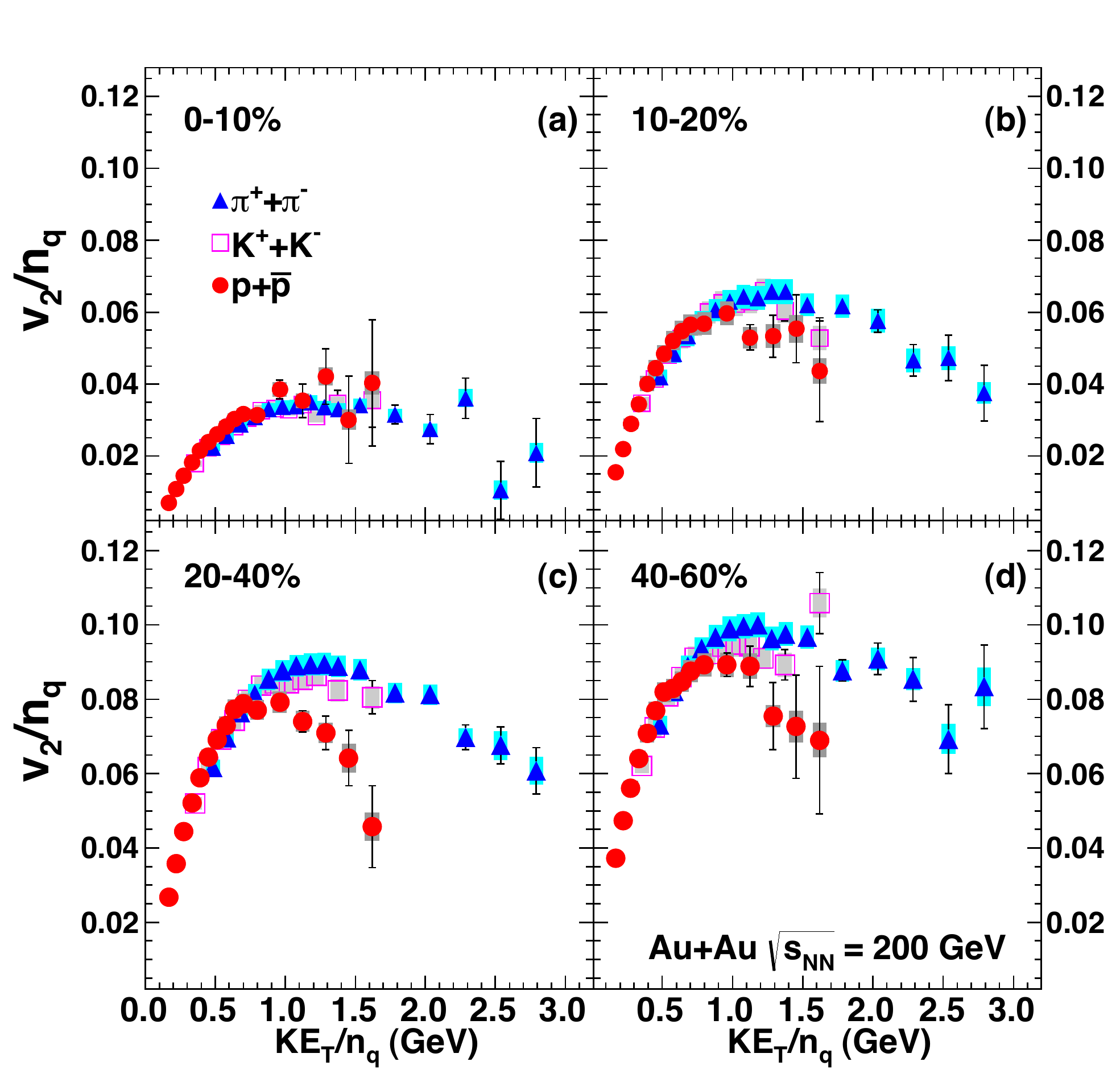}
  \caption{
  	\label{fig:nq_scaling}
  	PHENIX identified particle $v_2/n_q(KE_T/n_q)$ from $\sqrt{s_{NN}}$ = 200 GeV Au+Au collisions for centralities of ($a$) 0-10\%, ($b$) 10-20\%, ($c$) 20-40\%, and ($d$) 40-60\%~\cite{bib:v2_pid}.  The identified particles measured are pions (triangles), kaons (squares), and protons (circles).  Statistical and systematic errors are shown by the bars and boxes, respectively.  For 0-10\% centrality $n_q$ scaling holds throughout the measured range, but breaks at high $KE_T$ for the more peripheral collisions.  }
\end{figure}

Previous flow measurements~\cite{bib:Lacey} have shown that identified particle $v_2$ signals scale at low transverse kinetic energy ($KE_T$) when their quark constituent numbers are taken into account.  In other words, scaling is seen when plotting $v_2/n_q(KE_T/n_q)$, where $n_q$ is the number of quark constituents in the particle.  This provides strong evidence for quark degrees of freedom.  Shown in Fig.~\ref{fig:v2_pT_pid} are recent PHENIX results~\cite{bib:v2_pid} displaying $v_2(p_T)$ of pions, kaons, and protons to a higher $p_T$ than previously measured~\cite{bib:v2_pid_old}.  In ($a$),  0-20\% centrality, a potential plateau is seen at high $p_T$ for the protons that is not seen in ($b$), 20-60\% centrality.  Figure~\ref{fig:nq_scaling} shows $v_2/n_q(KE_T/n_q)$ for centrality ranges of 0-10\%, 10-20\%, 20-40\%, and 40-60\%.  For all the centrality ranges, $n_q$ scaling is seen at low $KE_T/n_q$.  However, only for the most central collisions (0-10\%) does the scaling hold for the entire measured range.  The scaling breaks in 10-20\% centrality collisions at $KE_T/n_q\approx1.1$ GeV and differences increase with decreasing centrality.  This behavior demonstrates a clear centrality dependence for identified particle $v_2$ scaling.


\section{Higher Order Harmonic Flow ($v_3$, $v_4$)}

\begin{figure}
	\centering
	\includegraphics[width=0.75\textwidth]{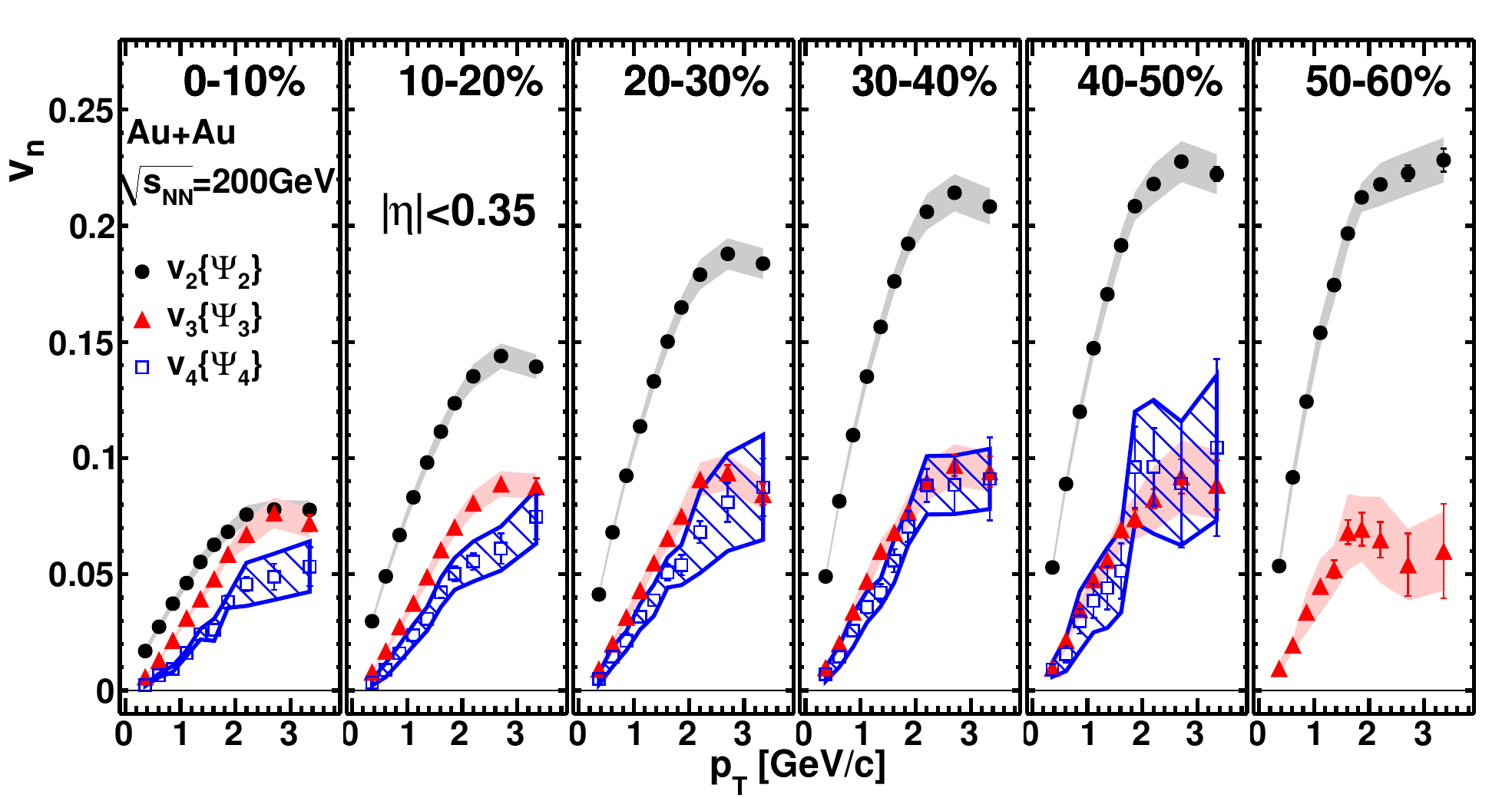}
	\caption{ 
		\label{fig:vn_pT_200GeV}
		PHENIX charged particle $v_n(p_T)$ from $\sqrt{s_{NN}}$ = 200 GeV Au+Au collisions for different centrality ranges within $|\eta|<$ 0.35~\cite{bib:PHENIX_vn}.  Here, $n$ has the same value in both $v_n$ and $\Psi_n$, with n = 2 (circles), n = 3 (triangles), and n = 4 (squares).  Statistical and systematic errors are shown by the bars and bands, respectively. }
\end{figure}

Higher order harmonic flow measurements, such as $v_3$ and $v_4$, have recently started revealing insights about the created medium's initial geometry.  Traditionally, the medium has been thought of as having a smooth Woods-Saxon matter distribution, in which case the odd harmonics ($v_1$, $v_3$, $v_5$...) would cancel out due to symmetry at pseudorapidity ($\eta) \approx 0$ and be compatible with zero.  However, recent studies~\cite{bib:triangular_flow} have shown that this may not be the case due to fluctuations in the participant geometry, causing the flow signal to persist for odd harmonics.  Figure~\ref{fig:vn_pT_200GeV} shows that indeed the $v_3$ signal is significantly positive near $\eta\sim0$ with a weak centrality dependence.  Both of these observations are consistent with initial geometry fluctuations.  

\begin{figure}
	\centering
	\includegraphics[width=0.55\textwidth]{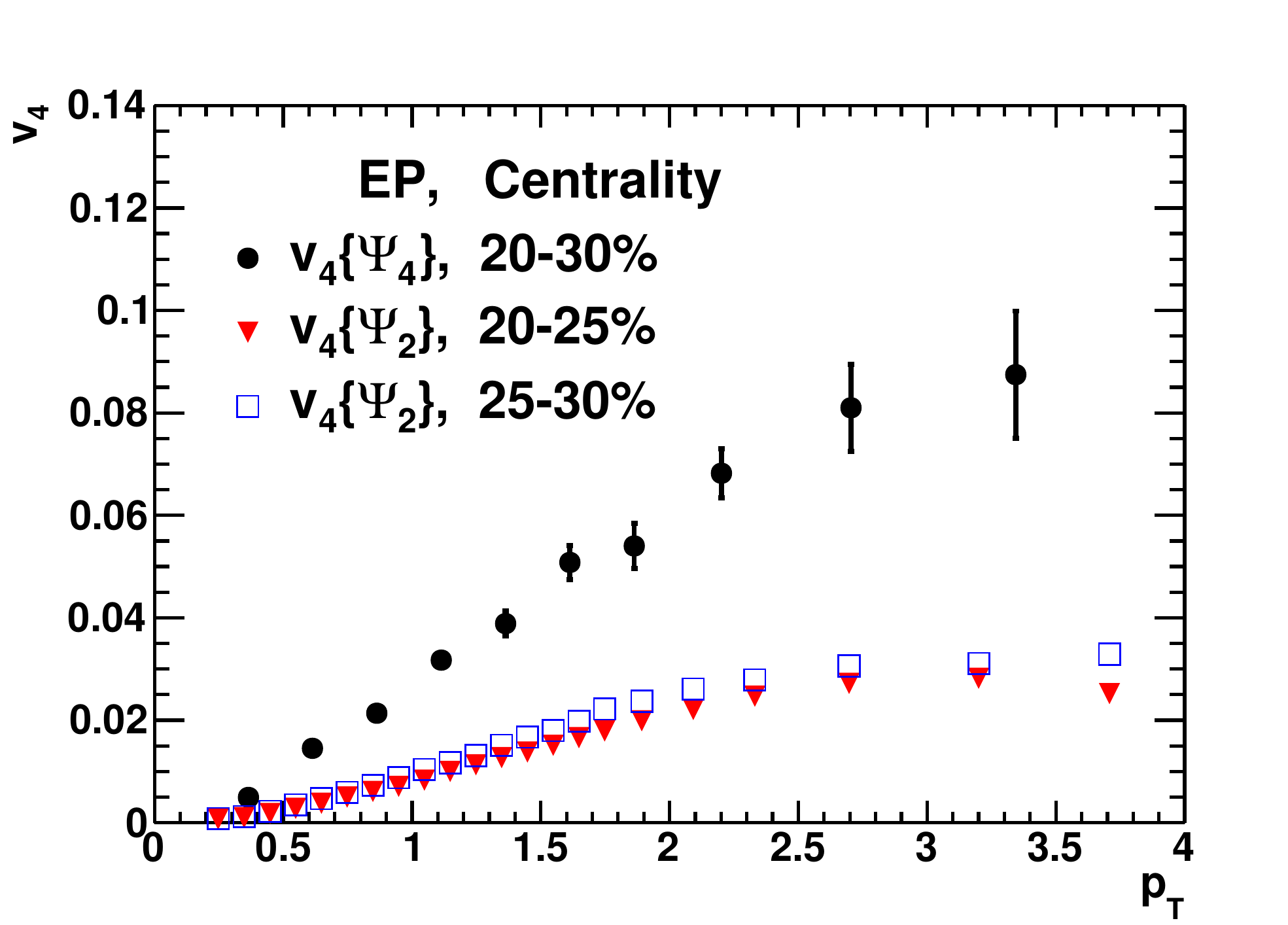}
	\caption{ 
		\label{fig:v4_pT_diff_harm_EP}
		Comparison of similar centrality charged particle $v_4(p_T)$ measurements using different harmonic event planes (EP).  The event plane and centrality combinations shown are: $\Psi_4$ and 20-30\% (circles), $\Psi_2$ and 20-25\% (triangles), and $\Psi_2$ and 25-30\% (squares).  For clarity only statistical errors are shown.  All data is from $\sqrt{s_{NN}}$ = 200 GeV Au+Au collisions using the PHENIX detector~\cite{bib:PHENIX_vn,bib:ppg098}. }
\end{figure}

Additionally, Fig.~\ref{fig:v4_pT_diff_harm_EP} shows that $v_4$ measured using $\Psi_4$ is $\sim$2x larger than when using $\Psi_2$ due to $\Psi_4$ originating not only from the same eccentricity and pressure gradients that drive $\Psi_2$, but also the same fluctuations as $v_4$.  This results in a stronger correlation and larger signal.  These fluctuations are thought to originate from the same type of initial geometry or eccentricity fluctuations that drive $v_3$.   Figure~\ref{fig:v4_pT_diff_harm_EP} emphasizes that eccentricity fluctuations are an important component in $v_4$ and need to be taken into account.  

\section{RHIC Beam Energy Scan}

In addition to ongoing top energy collisions, RHIC is also undertaking a staged multi-year beam energy scan, meaning colliding Au nuclei at energies less than its top energy of $\sqrt{s_{NN}}$ = 200 GeV.  The beam energy scan is valuable for many reasons, including studying how initial geometry effects change with beam energy, and probing the nuclear phase diagram in an attempt to find the critical point.  For these and other important energy scan measurements, flow is an ideal tool for investigation.  

\begin{figure}
	\centering
	\includegraphics[width=0.65\textwidth]{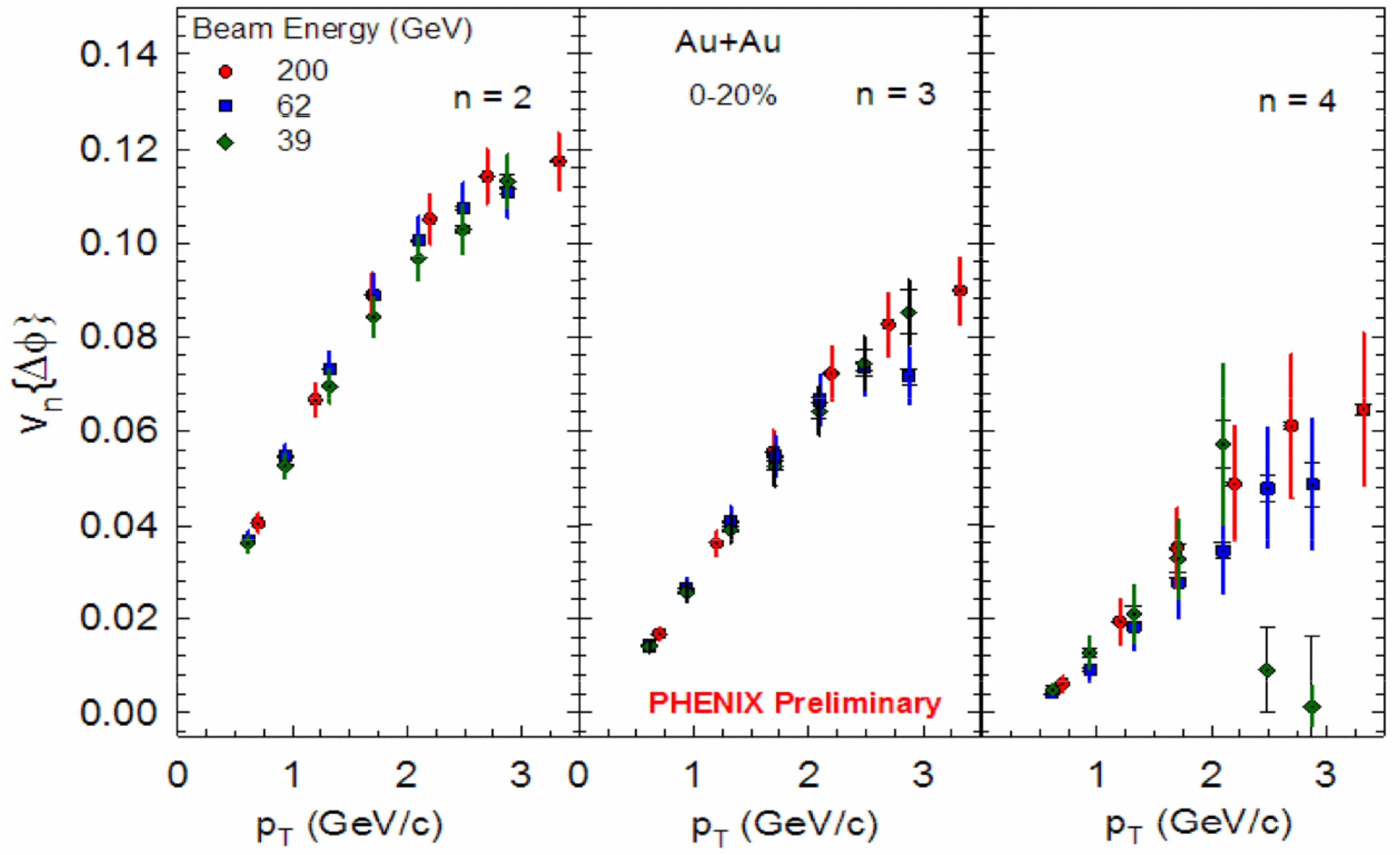}
	\caption{ 
		\label{fig:vn_pT_BES}
		PHENIX preliminary charged particle $v_n(p_T)$ for $\sqrt{s_{NN}}$ = 200, 62.4 and 39 GeV beam energies using a centrality range of 0-20\%~\cite{bib:Esumi_QM_talk}.  Statistical and systematic errors are shown by the bars and colored lines, respectively.  Despite the significant differences in beam energies the flow signals are consistent within the different harmonics, indicating similar hydrodynamic properties and initial geometry effects.  The measurement was done using a two particle correlation method between mid and forward rapidities. }
\end{figure}

For instance, one might expect that approaching the critical point would result in changes to the medium's behavior, such as a change in flow signal.  
Figure~\ref{fig:vn_pT_BES} compares RHIC's top energy $\sqrt{s_{NN}}$ = 200 GeV $v_n(p_T)$ data to lower beam energies of 62.4 and 39 GeV for flow harmonics of $n$ = 2, 3, 4.  Despite a factor as large as $\sim$5 difference in beam energy, the different energy flow signals are consistent within errors, meaning that within this energy range the hydrodynamic properties of the medium, as well as initial geometry effects, are similar.  

\begin{figure}
	\centering
	\includegraphics[width=0.77\textwidth]{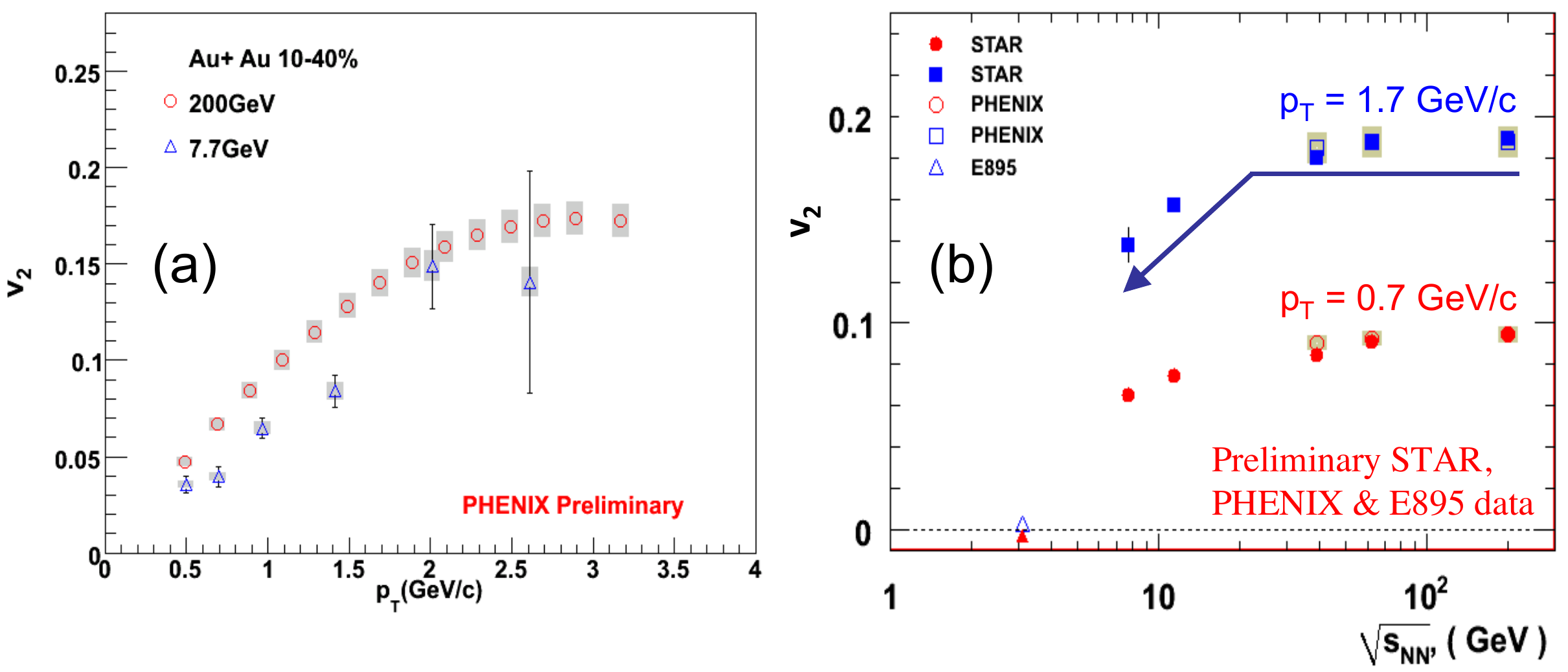}
	\caption{ 
		\label{fig:v2_pT_7GeV}
		($a$) PHENIX preliminary data for charged particle $v_2(p_T)$ at $\sqrt{s_{NN}}$ = 7 GeV (triangles) and 200 GeV (circles)~\cite{bib:shengli_winter_workshop}.  The difference in $v_2$ indicates a change in the medium's properties.  ($b$) Charged particle $v_2(\sqrt{s_{NN}})$ at specific $p_T$ values for various beam energies from PHENIX, STAR, and E895~\cite{bib:Esumi_QM_talk}.  Here a flattening of the signal is seen above $\sqrt{s_{NN}}$ = 39 GeV, indicating similar medium properties, while a decrease in signal begins in a transition region somewhere below this energy, indicating a change in the medium's properties. }
\end{figure}

Conversely, if the $v_2(p_T)$ result from $\sqrt{s_{NN}}$ = 7 GeV is compared to that from 200 GeV, as shown in Fig.~\ref{fig:v2_pT_7GeV}($a$), a significant difference is seen, indicating a change in the medium's properties.  Displayed another way, Fig.~\ref{fig:v2_pT_7GeV}($b$) shows $v_2(\sqrt{s_{NN}})$ at $p_T$ = 0.7 and 1.7 GeV/c, where the signal flattens between 39 and 200 GeV, but starts decreasing somewhere below 39 GeV.  Does this indicate the medium is changing from partonic to hadronic interactions?  Can this help reveal the critical point?  Investigations are continuing in this transition region which will help in answering these questions, including the collection of new data at $\sqrt{s_{NN}}$ = 19.6 and 27 GeV during the 2011 RHIC data taking period.

\section{Summary}

Recent PHENIX results show that the $n_q$ scaling seen in previous measurements breaks at larger $KE_T$ values and has a significant centrality dependence.  Additionally, higher order harmonic flow measurements provide insights into the medium and can be specifically applied to constrain initial geometric fluctuations.  Lastly, RHIC's beam energy scan is helping to understand the hadron gas and quark gluon plasma phase transition, as well as constrain the location of the critical point.  More PHENIX beam energy scan results can be found in A. Franz's proceedings in this volume.  The PHENIX collaboration continues working in all of these areas in hopes of making additional progress via ongoing detector upgrades and the collection of new data sets.

\end{document}